# Mode-Locked Fiber Laser with up to 19 kHz Wavelength Sweep Rate via External Pump LD Modulation

Guanyu Ye[1], Maolin Dai[1], Bowen Liu[1], Yifan Ma[1], Takuma Shirahata[2], Shinji Yamashita[1,2] and Sze Yun Set [1,2]*

[1]*Department of Electrical Engineering and Information Systems, The University of Tokyo. Bunkyo-ku, Tokyo 113-8656, Japan*

[2]*Research Center for Advanced Science and Technology, The University of Tokyo. Meguro-ku, Tokyo 153-8904, Japan*

E-mail: set@cntp.t.u-tokyo.ac.jp

For the first time, we introduce a rapid wavelength-swept, passively mode-locked fiber laser in an all-polarization-maintaining and all-fiber configuration. Achieving an exceptional wavelength sweep rate of up to 19 kHz through external modulation of the LD driver pump current, this laser offers a high sweep rate, simple cavity design, cost-effectiveness, and excellent repeatability.

Wavelength-tunable ultrashort mode-locked fiber lasers (MLFLs) hold great promise for applications in spectroscopy [1], optical sensing [2], and the biomedical field [3]. Among these lasers, tunable models capable of sweeping pulse wavelengths at rates of hundreds of hertz or more are beneficial in various advanced imaging techniques. These include stimulated Raman scattering (SRS) microscopy for non-invasive, label-free, real-time observation of molecules and cells [4, 5]; two-photon excitation (2PE) microscopy for reduced photodamage and deeper tissue imaging [6, 7]; and difference frequency generation (DFG) for mid-infrared (MIR) spectroscopy [8].

Various methods for achieving wavelength tunability include temperature or strain adjustments of intracavity spectral filters [9], manual birefringence control via polarization controllers (PCs) [10], and adjusting intracavity loss using variable optical attenuators (VOAs) [11]. However, non-polarization maintaining (PM) configurations in these studies can adversely affect the stability and reliability of laser performance. Moreover, the tuning methods used in these lasers have inherent drawbacks for high-speed wavelength tuning: manual adjustments are difficult to replicate consistently, while thermal tuning responds slowly as temperature changes need time to take effect. Techniques such as dispersion-tuned mode-locking (DTML) and Fourier domain mode-locking (FDML) can achieve wavelength sweeping rates exceeding kHz levels [12, 13]. However, these methods necessitate intricate dispersion management, complicating the laser system. They employ complex and costly active mode-locking mechanisms that require externally driven modulators and advanced modulation control. Additionally, the generated pulses are typically in the nanosecond range or longer. Such quasi-continuous-wave light with low peak power suits applications like optical coherence tomography (OCT) [14]. However, it is less effective for applications requiring ultrashort pulses, such as SRS or MIR generation.

Compared to active wavelength-swept sources, a passively MLFL configured with all-polarization-maintaining (all-PM) components and employing a saturable absorber offers a more straightforward design. This configuration produces femtosecond or picosecond ultrashort pulses with enhanced reliability and stability, making it suitable for applications such as SRS [15], MIR generation [16], or 2PE microscopy [17]. In 2015, Ozeki et al. introduced an all-PM MLFL featuring a galvanometer scanner (GS) and diffraction grating, achieving wavelength sweeping by adjusting the incident angle on the grating through the rotation of the GS [18]. This setup employs a semiconductor saturable absorber mirror for passively mode-locking, delivering picosecond ultrashort pulses with a sweep rate of up to 500 Hz. More recently, in 2023, Sun et al. developed a different approach utilizing a built-in intracavity Lyot filter made from PM fiber segments, angle splicing, and a polarizer [19]. The filter's free spectral range (FSR) is mechanically adjusted by stretching the PM fiber using a voltage-controlled piezoelectric ceramic (PZT). This carbon nanotube-based all-PM, all-fiber MLFL has achieved a sweep rate of up to 500 Hz. Limited by the mechanical responses of the GS and PZT, these studies could not achieve sweep rates beyond 500 Hz. Currently, there are no reports of an all-PM, all-fiber passively MLFL capable of delivering ultrashort wavelength-swept pulses at kHz-level sweep rates. In this study, we demonstrate an innovative approach to generating ultrashort pulses with a wavelength sweep rate of up to 19 kHz simply by externally modulating the pump current of the laser diode (LD) driver. Compared to existing wavelength-sweeping methods, this electrically controlled approach offers a more cost-effective and straightforward solution for generating high sweep rate, wavelength-swept ultrashort pulses.

Fig. 1 illustrates the fast wavelength-swept laser schematic diagram, comprising an external modulation section and a tunable laser part. The tunable laser is recently reported in our previous research [20], an all-PM all-fiber MLFL based on nonlinear polarization evolution (NPE). NPE mode-locking relies on a nonlinear phase relationship that depends on both the wavelength and pulse peak power. Therefore, changing the pump power alters the pulse peak power, consequently shifting the lasing wavelength to meet the phase criteria required for NPE mode-locking. A 20 nm tuning range (~1568 to 1589 nm) is achieved by varying the pump power. Compared with the thermal or mechanical tuning method, this electrically controllable pump power tuning technique enables rapid response, making it suitable for high-speed wavelength sweeping. The external modulation section includes a signal generator (Agilent 3320A) and a pump LD driver (Thorlabs CLD1015). The LD driver is configured to "External Modulation" mode, which allows the pump current to be adjusted via an external voltage. The signal generator produces sinusoidal waveforms at designated frequencies and voltages fed into the LD driver. This driver then outputs a modulated pump current to the pump LD, effectively modulating the pump power and facilitating rapid wavelength sweeping. An optical spectrum analyzer (OSA, YOKOGAWA AQ6370D) is configured in 'repeat sweep' mode with the 'Max hold' setting to capture the center wavelength trace, representing the sweep range achieved throughout the wavelength sweeping experiment.

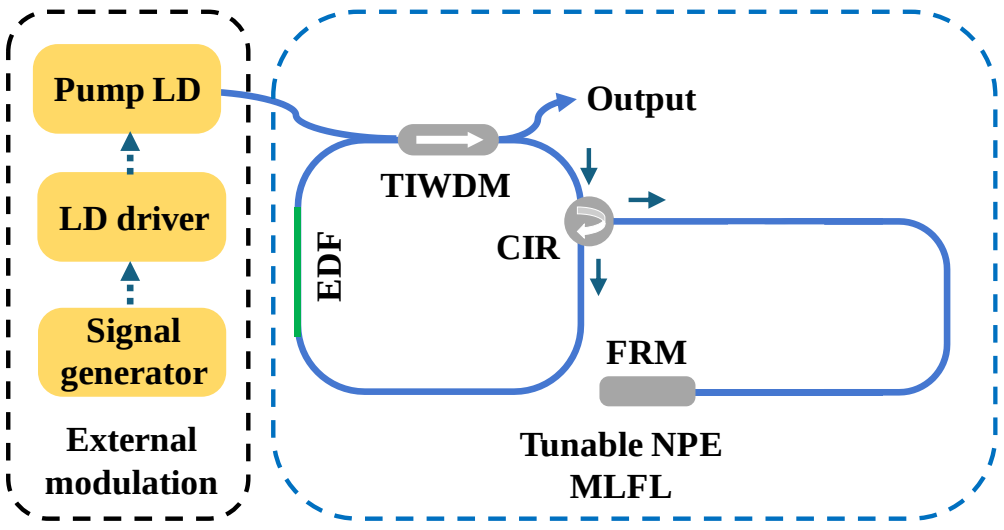

Fig. 1. Schematic diagram of the high-speed wavelength-swept MLFL. All fibers and components in the tunable NPE MLFL are PM type. TIWDM: PM tap isolating wavelength-division multiplexing, FRM: Faraday rotation mirror, CIR: circulator, EDF: Erbium-doped fiber.

The Max-hold spectra at lower sweep frequencies of 100 Hz, 200 Hz, and 500 Hz are displayed in Fig. 2(a), 2(b), and 2(c). At a 100 Hz sweep rate, the spectrum spans 20 nm, closely matching the 20.7 nm achieved with manual tuning, suggesting minimal shortening of the tuning range at frequencies up to this level. The tuning range at 200 Hz is determined by comparing the max-hold spectra at 200 Hz with that at 100 Hz. Upon superimposing these spectra, an overlapping region is identified. This overlapping segment represents the effective tuning range at 200 Hz. Similarly, the tuning ranges at other sweep rates are analyzed using this method. However, the tuning range decreases to 14 nm at 200 Hz, with the shortest and longest wavelengths shifting toward the central wavelength. At 500 Hz, the range narrows further to 7.5 nm, continuing the convergence trend toward the central wavelength. Fig. 2(d), 2(e), and 2(f) display the tuning range at higher frequencies of 1 kHz, 10 kHz, and 19 kHz. The tuning cannot exceed 19 kHz due to the LD driver's maximum external signal frequency tolerance of 20 kHz. At these frequencies, the tuning range diminishes to 4.2 nm, 3.8 nm, and 3.3 nm, respectively. This reduction follows a similar pattern to lower frequencies, with both the starting and stopping wavelengths converging towards the center wavelength. The laser consistently maintains mode-locking throughout the tuning process at these sweep rates, and the Max-hold spectrum stabilized within tens of seconds. Given the LD driver's maximum tolerance frequency of 20 kHz for external modulation, we cannot sweep the laser at higher rates.

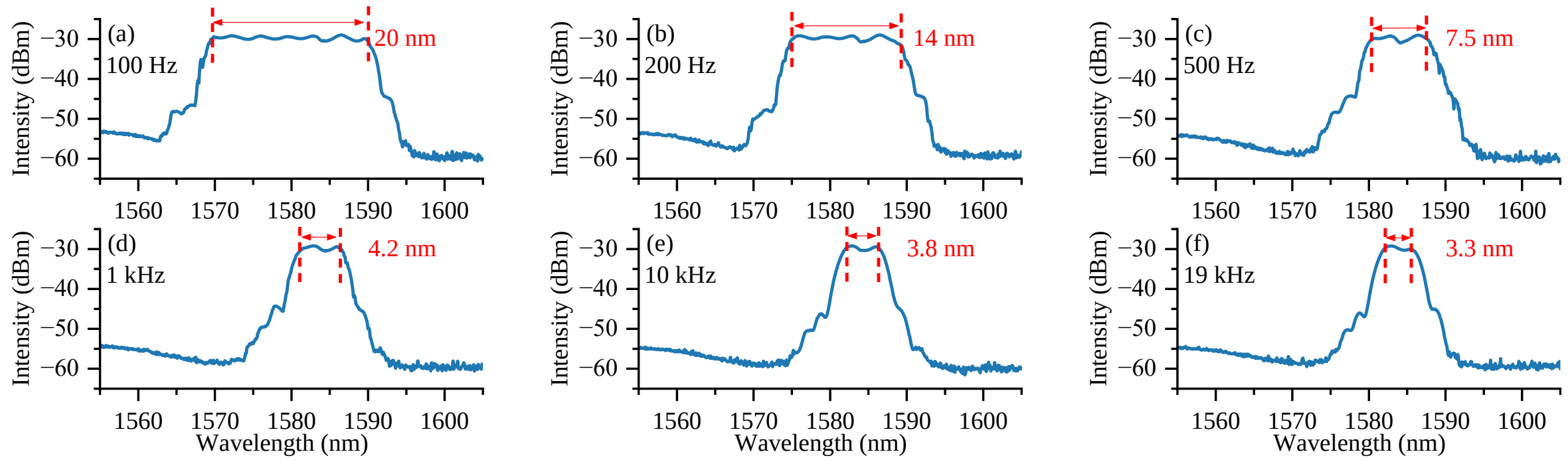

Fig. 2. Max-hold spectra at varying sweep rates. (a)-(c) show 100 Hz, 200 Hz, and 500 Hz; (d)-(f) show 1 kHz, 10 kHz, and 19 kHz, respectively.

The radio frequency (RF) spectrum is captured using a photodetector (New Focus 1611) and an electrical spectrum analyzer (ESA, RIGOL RSA3045). Fig. 3 (a) shows the RF spectrum under static conditions with no external modulation, where a high SNR of 75 dB indicates high pulse stability. Fig. 3(b) illustrates the RF spectrum when a 100 Hz external signal modulates the LD driver, and the broadening at the base of the RF peak suggests modulation of the center frequency. The results at sweep rates of 10 kHz and 19 kHz are displayed in Fig. 3(c) and Fig. 3(d). The RF spectra clearly show a central frequency peak flanked by two sidebands with lower amplitudes. These sidebands are symmetrically spaced around the central frequency, with frequency differences corresponding to the sweep rates: 10 kHz in Fig. 3(c) and 19 kHz in Fig. 3(d). These sidebands arise from frequency modulation induced by the external driving signal. Since the laser's repetition rate varies with wavelength, sweeping the wavelength also modulates the repetition rate. Consequently, in the RF spectrum, the external modulation signal shifts the center frequency to create symmetrical sidebands, with a separation corresponding to the modulation signal frequency [21]. Sideband characteristics, such as their number and intensity, depend on the modulation depth and frequency. The modulation depth is related to the variance in repetition rates between the tuning range's start and end wavelengths. A narrow tuning range of only a few nanometers leads to a slight modulation depth, producing only two noticeable sidebands on the RF spectrum, as shown in Fig. 3 (c) and 3(d). Due to the limitations of the ESA, the expected multiple symmetric sidebands, which should be spaced 100 Hz from the center frequency, are not discernible in Fig. 2(b). Instead, a broadened spectral base is observed, indicating that the repetition rate undergoes modulation by the external signal.

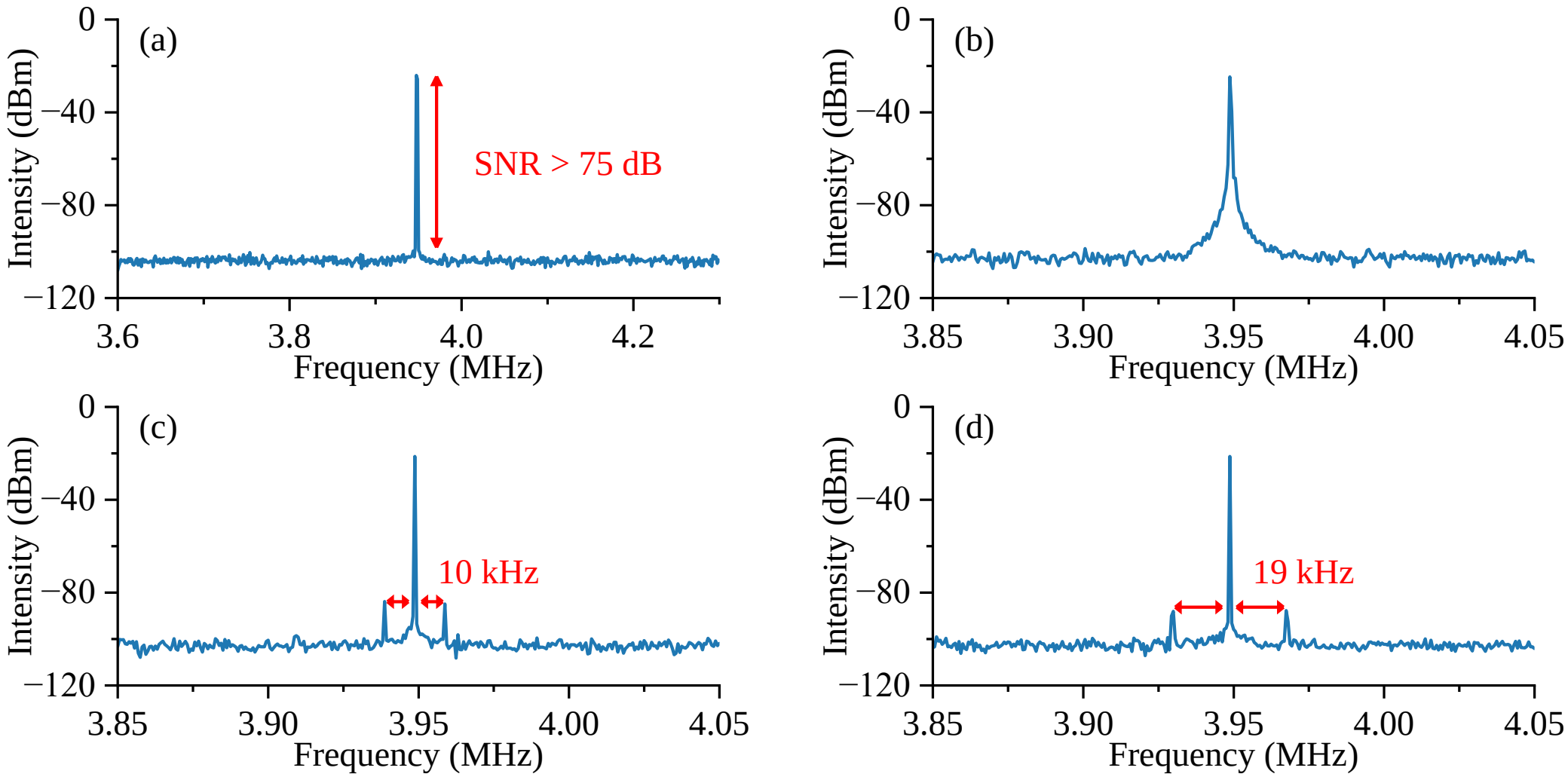

Fig. 3. RF spectra with a resolution bandwidth (RBW) of 100 Hz showing: (a) static without wavelength sweeping; (b) 100 Hz sweep rate; (c) 10 kHz sweep rate; (d) 19 kHz sweep rate.

The sweep rate of the center wavelength is influenced by the frequency of the modulation signal, with a noted reduction in tuning range at higher frequencies. At 200 Hz, the sweep range narrows to 70% of the range observed at 100 Hz. At 500 Hz, it further reduces to ~ 37.5%; at 19 kHz, the sweep range diminishes to ~ 16.5%. This reduction in sweep range at higher frequencies is consistent with findings from other passive wavelength-swept laser sources, validating the reliability of our results under comparable experimental conditions. In the all-PM fiber Lyot filter study [19], the tuning range decreases to 70% and 48% of its value at 100 Hz when increased to 300 Hz and 500 Hz, respectively. Similarly, in the research involving galvanometer scanners and gratings [18], the tuning range at 500 Hz is reduced to 40% of the range observed at 100 Hz.

The reduction in tuning range may arise from two main factors. The first possible reason is the mode-locking pulse build-up dynamics, where pulse stabilization and build-up require time following changes in cavity parameters [22, 23]. Thus, if the pump power sweeps faster than the pulse buildup process—where the pump condition changes before the new pulse stabilizes—then a stable pulse at specific wavelengths may not be achieved during the sweep experiment. Second, the response of the LD driver might limit the effective scanning of the pump power at high modulation frequencies, even though it can handle frequencies up to 20 kHz. Since the pump power adjusts the center wavelength, a limited adjustable range of pump power at high frequencies naturally results in a reduced wavelength tuning range. Fig. 4(a) illustrates the schematic setup for measuring the LD pump power sweep range. The signal generator delivers an undistorted sinusoidal signal at various frequencies, maintaining consistent amplitude, to modulate the LD driver's output pump current. The LD output power is modulated and directed through an attenuator to protect the photodiode (PD). An oscilloscope captures the electrical signal from the PD, and the peak-to-peak voltage (Vpp) of the oscilloscope trace indicates the pump power sweep range. Fig. 4(b) to 4(e) display the oscilloscope traces corresponding to signal generator frequencies set at 100 Hz, 1 kHz, 10 kHz, and 19 kHz. The oscilloscope is configured to the same amplitude scale to facilitate a clear comparison of the pump power sweep results across different frequencies. These traces exhibit a sinusoidal waveform with a stable maximum and minimum amplitude over time, evidenced by a flat envelope at the peaks and valleys. This stability validates the effectiveness of sweeping the pump power via external modulation of the LD driver, confirming its reliability at high frequencies. While the reduction in pump power sweep range (represented by Vpp) is minimal at 1 kHz in Fig. 4(b), it decreases drastically at higher frequencies of 10 kHz and 19 kHz, as evident in Fig. 4(c) and 4(d).

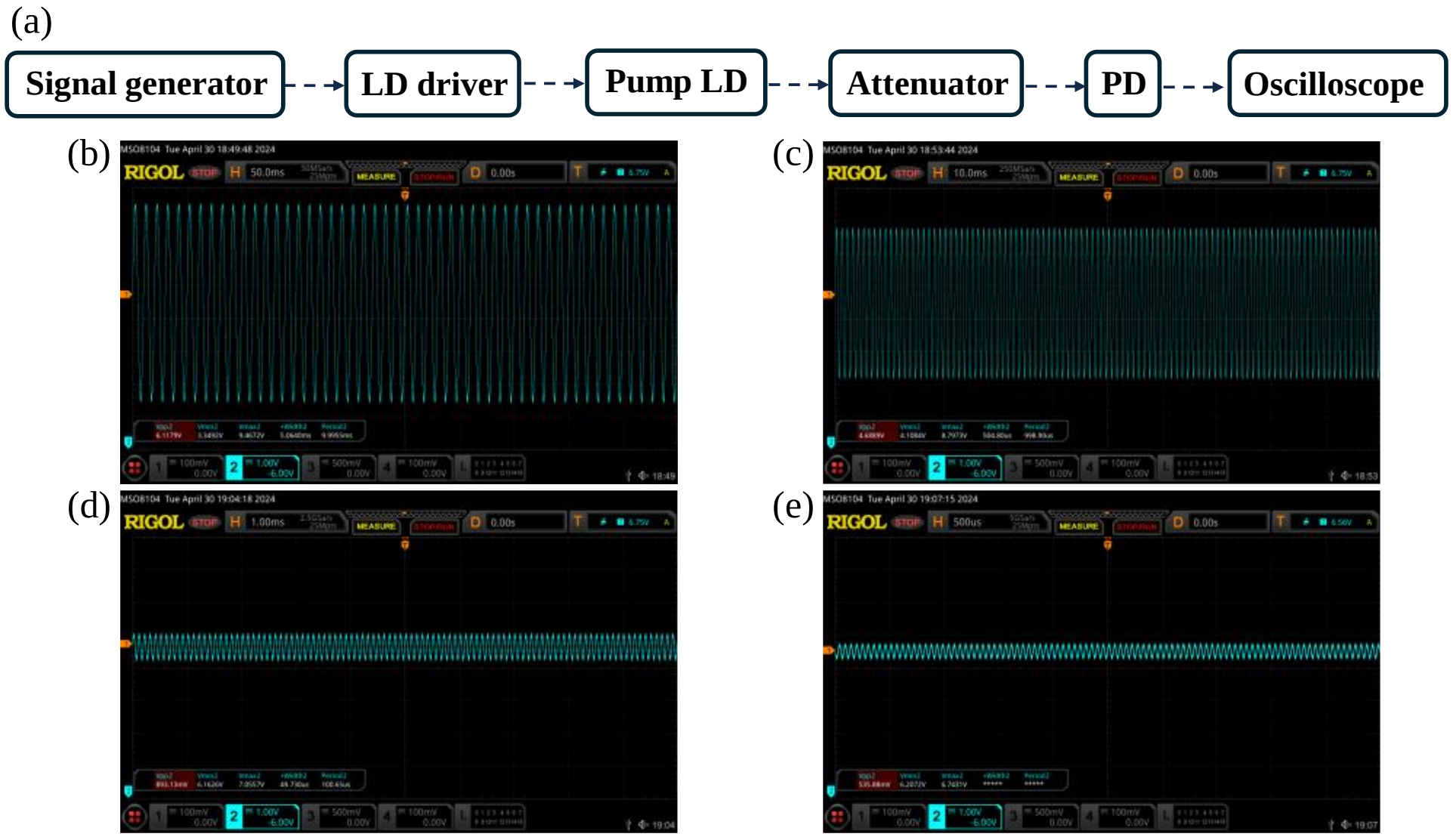

Fig. 4. (a) Schematic setup diagram for measuring LD pump power sweep range at various sweep rates. (b)-(e) Oscilloscope traces at sweep rates of 100 Hz, 1 kHz, 10 kHz, and 19 kHz, respectively.

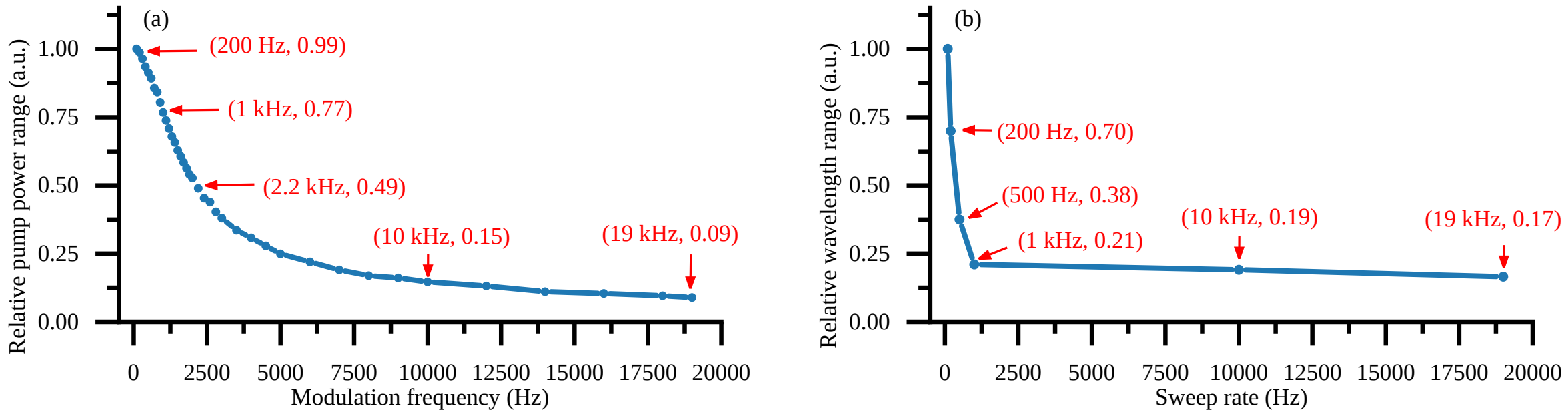

Fig. 5 Relative sweep ranges at various modulation frequencies: (a) pump power and (b) wavelength sweeping.

Fig. 5(a) illustrates the relative pump power sweep range (represented by Vpp of the oscilloscope trace) at various modulation frequencies. We use the sweep range at 100 Hz as a baseline for measuring and calculating the relative sweep range at higher frequencies. It is observed that the relative range consistently decreases with increasing modulation frequency. It diminishes to half at a 2.2 kHz rate and drops below 15% at frequencies exceeding 10 kHz. In our laser system, where wavelength sweeping is achieved through pump power adjustments, this reduction in power sweep range consequently restricts the overall wavelength sweep range achievable in mode-locking, particularly at kHz-level sweep rates.

Fig. 5(b) displays the relative wavelength-swept range at various sweep rates, using a 20 nm sweep at 100 Hz as the baseline. At 200 Hz, while the wavelength tuning range only achieves 70% of that at 100 Hz in Fig. 5(b), the LD pump power sweeps 99% of the range at 100 Hz in Fig. 5(a). This phenomenon, where the wavelength sweep range decreases despite the pump LD covering almost the entire adjustable power range, indicates that the primary cause of the reduction in wavelength sweep range is the pulse buildup dynamics. A similar situation occurs at a sweep rate of 500 Hz, where the pump LD covers 91% of its power range, but the wavelength sweep is limited to only 38% of its total range. At 1 kHz, the pump power covers 77% of its capacity, while the wavelength sweep covers 21% of its total range. This indicates that although the LD response starts to contribute, the pulse buildup dynamics remain the main factor in the wavelength sweep reduction. Beyond 1 kHz, the wavelength tuning range shortens gradually. At sweep rates of 10 kHz and 19 kHz, the LD response becomes the primary factor in reducing the wavelength sweep range, with pump power covering only 15% and 9% of its capacity, respectively. Due to the nonlinear response of wavelength shifting to pump power changes [20], there is ~17% of the wavelength sweep range at a 19 kHz sweep rate.

Several possible approaches enhance the performance of this wavelength-swept passive MLFL. Improvements can be achieved by enlarging the maximum achievable wavelength tuning range, shortening the pulse buildup time, and strengthening the pump LD response. First, the gain medium and tuning mechanism limits the achievable wavelength tuning range, where the pulse center wavelength redshifts with increasing pump power, contrary to the gain curve shift [24]. This opposing trend limits the longest and shortest tunable wavelengths, constraining the achievable tuning range. To address this, the tuning range can be expanded by modifying the gain profile or using a gain medium with a broader gain spectrum [25]. Second, since the pulse buildup time is influenced by the cavity round-trip time, pump power, and cavity losses [22], modifying the cavity parameters—such as reducing the overall cavity length, lowering intracavity losses, and increasing pump power—can decrease the time required for pulse buildup. Third, because the LD driver cannot thoroughly sweep the entire pump power range at high frequencies, a more sophisticated LD driver and pump LD can enhance its response at these frequencies.

In conclusion, we present the first wavelength-swept passive MLFL capable of achieving a sweep rate of up to 19 kHz. Our non-mechanical, electrically controllable approach improves the sweep rate 38 times compared to existing methods. This rapid pulse wavelength sweeping is achieved by fast modulation of the pump power through external modulation of the LD driver pump current. Our results confirm the feasibility and potential of this high-speed wavelength sweep method. The simple configuration of our all-PM, all-fiber laser design offers a practical, cost-effective, reliable, and user-friendly solution for rapid wavelength sweeping, demonstrating considerable promise for various applications.

**Acknowledgments**

This research is supported by the Japan Society for the Promotion of Science (18H05238, 22H00209, 23H00174) and Core Research for Evolutional Science and Technology (JPMJCR1872). Sze Yun Set thanks Mr. Hideru Sato for the personal donation partially supporting this research.